\documentstyle[12pt]{article}
\begin{document}
\baselineskip=15truept
\font\tyt=cmcsc10 scaled\magstep2
\font\tlu=cmssbx10 scaled\magstep3
\centerline{\tlu Evaluation of Bartnik's} 
\medskip
\centerline{\tlu quasilocal mass function
\footnote{This paper is supported by the Polish Government Grant no 
2PO3B 090 08.}}
\vskip 2.5cm
\centerline{\bf Piotr Koc}
\vskip 1.5cm
\centerline{\bf Institute of Physics
\footnote{\rm e-mail address: \tt ufkoc@ztc386a.if.uj.edu.pl}}
\centerline{\bf Jagellonian University}
\centerline{\bf Reymonta 4, 30-059 Krak\'ow,}
\centerline{\bf Poland}
\vskip 2.5cm
\centerline{PACS number: 04.20.Cv}
\vskip 1.5cm
\centerline{\bf Abstract}
Bartnik's definition of gravitational quasilocal energy 
is analyzed. For a wide class of systems Bartnik's function is
given by the ADM mass of some vacuous extension. 
As an example we calculate mass of a non central 
ball in Schwarzschild 
geometry. The ratio mass to volume becomes 
singular in the limit of 
small volumes.
\vfill
\eject

\par
One of the still unsolved issues of 
classical gravitation is a problem of 
the quasilocal measure of gravitational energy. 
Although from the formal point
of view there is no need to introduce 
such quantity in a framework of
general relativity there has been made 
a great effort \cite{enpola}\cite{bartnik} 
to construct mass function 
which interpolates between the local energy - 
momentum tensor and the 
global ADM mass \cite{ADM}. One of possible motivations of 
these investigations
follows from the desire to find tools which would 
support the formulation of 
conditions of the gravitational collapse \cite{cond}. 
Difficulties in the construction
of theorems which diagnose the existence of horizons 
are mainly connected 
with the absence of concepts of a quasilocal 
mass and a size of the configuration.
\par
It seems that until now the best candidate for the definition 
of quasilocal energy
is a proposal of Bartnik \cite{bartnik}.
In the case of a time-symmetric hypersurface Bartnik's 
mass is defined as 
follows. Let us consider the class of Riemannian three manifolds:
\par
\bigskip
\bigskip
${\cal {PM}}=\{(M,g): (M,g)$ 
{\it is an asymptotically flat time-symmetric
initial data set, satisfying the weak 
energy condition and such $M$ has no
horizons}\}
\par
\medskip
\bigskip
\bigskip 
Mass of a domain $\Omega$ (with $\partial \Omega$ connected) 
is defined as the infimum of ADM masses of all 
possible (physical) {\bf extensions} of $\Omega$.
$$m_B(\Omega)=
inf\{m_{ADM}(\widetilde M):
\Omega \subset\widetilde M\in{\cal {PM}}\}$$
\par
The no-horizon condition was imposed 
originally by Bartnik to exclude explicit
examples of solutions constructed in such a way that the ADM mass
becomes arbitrary small. Physically 
these solutions correspond to systems 
which are formed by adding a matter 
inside the horizon. Such operation 
decreases both the area of the horizon and the ADM mass. 
\par
In this paper we prove that the Bartnik's quasilocal energy of
conformally flat domain $\Omega$ is given by 
the ADM mass of some $C^0$ extension $\widetilde M_{cfv}$ 
which is conformally flat and {\bf vacuous outside} $\Omega$. 
Therefore in a wide class of applications it is sufficient to 
analyze only vacuous extensions.  This conjecture was originally 
formulated in \cite{bartnik}. Another issue is  to choose the 
specific vacuous conformally flat extension which does 
not contain horizons and minimalizes 
the ADM mass. We solve this problem on an explicit example 
in the Schwarzschild geometry. We calculate the quasilocal energy 
of gravitational field enclosed in a non central small ball which 
is a part of the spacelike hypersurface in the 
Schwarzschild solution. 
However the presented technique is 
entirely general and can be implemented 
for any conformally flat subset $\Omega$ of arbitrary geometry.
\par
In the below two theorems there is 
considered wider then $\cal{PM}$ class
of extensions in which the presence of horizons is allowed:
\par
\bigskip
\bigskip
$\cal{PMW}=\{(M,g): (M,g)$ {\it is an 
asymptotically flat time-symmetric
initial data set, satisfying the weak energy condition.}\}
\par
\medskip
\bigskip
\bigskip
We temporarily skip the problem of horizons and show 
that in $\cal{PMW}$ for any
"massive" extension 
$\widetilde M(\mu)$ ($\mu$ denotes matter density) 
there exist some vacuous 
(outside $\Omega$) extension
$\widetilde M(\mu=0)$ which has the ADM mass is not greater 
then mass of $\widetilde M(\mu)$.
Therefore evaluation of Bartnik's mass function
reduces to comparing ADM masses 
of vacuous extensions which have no horizons. 
In the case of conformal flatness different 
vacuous extensions are parameterized 
by the value of conformal factor 
at infinity. Finally on the example of
Schwarzschild geometry we will 
present the method of comparing the ADM 
masses of extensions with different asymptotics.
\par
As it was suggested by 
Bartnik the original definition may be 
specialized by the assumption of some 
additional properties of $\Omega$
and ${\cal {PM}}$ (axial or spherical symmetry, some curvature 
conditions etc.). In the following theorem we 
assume such kind of specialization.
\par
{\bf Theorem 1.} Assume conformall flatness 
$\Omega$ and $\cal{PMW}$. For any 
extension $\widetilde M(\mu_1)$ there exist some 
vacuous extension $\widetilde M(\mu=0)_{cfv}$ which 
has the ADM mass not greater then 
ADM mass of $\widetilde M(\mu_1)$: $m_1\geq m_{cfv}$.
\par
{\bf Proof.} Because of conformal flatness three 
dimensional line element
may be written as:
$$g_{ik}=\Phi^4 h_{ik},\eqno(1)$$
where $h_{ik}$ denotes cartesian flat metric.
The constraint equations $G_{0\mu}=0$ on a 
time-symmetric slice (i.e. $K_{ij}=0$) 
reduce to the hamiltonian constraint:
$$\nabla^2\Phi=-2\pi\mu\Phi^5,\eqno(2)$$
The weak energy condition implies 
that matter density $\mu$ is nonnegative;
$\mu\geq 0$. Asymptotically solution 
is flat: $\Phi\rightarrow A+{B\over 2r}+O(1/r^2)$, 
where positive constants $A$ 
and $B$ define the ADM mass of the system:
$$m_{ADM}=AB.\eqno(3)$$
Conformal factor $\Phi$ is uniquely determined 
by matter distribution $\mu$, 
constant $A$ and the boundary condition on $\partial \Omega$. 
Therefore we designate different 
extensions by specifying $\mu$ and  $A$: 
$\widetilde M(\mu,A)$ (boundary condition on $\partial\Omega$ is 
the same for all extensions).
Let us study an arbitrary conformally flat 
extension $\widetilde M(\mu_1, A_1)\in\cal{PMW}$. 
We will construct a conformally flat vacuous extension 
$\widetilde M(\mu=0,A_2)_{cfv}\in\cal{PMW}$ 
which has the ADM mass not greater then ADM
mass of $\widetilde M(\mu_1,A_1)$. 
For $\widetilde M(\mu_1, A_1)$ we have
$$\nabla^2\Phi_1=-2\pi\mu_1\Phi_1^5,\eqno(4)$$
For conformally flat vacuous 
extension $\widetilde M(\mu=0,A_2)_{cfv}$
constraints reduce to 
$$\nabla^2\Phi_{cfv}=0.\eqno(5)$$
By the definition the geometry of $\Omega$ 
is fixed but we have a freedom of determining
the constant $A_2$ at infinity. Let us choose $A_2=A_1$. 
Thus the difference $$\chi=\Phi_1-\Phi_{cfv}\eqno(6)$$
must vanish on $\partial \Omega$ 
and at infinity. Subtracting equations (4)
and (5) we get
$$\nabla^2\chi=-2\pi\mu_1\Phi_1^5.$$
Because of the nonnegativity of matter density 
function $\chi$ is superharmonic:
$\nabla^2\chi\leq0$. From min-max 
principle \cite{gilparg} we know that 
superharmonic function attains its infimum on 
the boundary (in our case on $\partial \Omega$
or at infinity). On the 
boundary $\chi$ equals zero therefore in domain $R^3-\Omega$
function $\chi$ has to be nonnegative and $\Phi_1\geq\Phi_{cfv}$. 
Using the fact that the constant $A$ is 
the same for both $\Phi_1$ and $\Phi_{cfv}$ 
we get $B_1\geq B_{cfv}$ and from (3):
$$m_1\geq m_{cfv}.$$
What ends the proof.
\par
We suppose that the statement of 
the theorem is also true for more 
general geometries. Below we study a class 
of extensions which is in some of 
sense "orthogonal" to the case of conformal flatness.
\par
Assume that $\Omega$ is conformally flat and let 
us restrict to axially symmetric $\Omega$ and $\cal{PMW}$
(we {\bf do not assume conformall flatness of} ${\cal {PMW}}$).
General form of the axially symmetric 
three dimensional line element is 
following \cite{brill}:
$$ds^2=\Phi^4(r,\vartheta)\bigl(e^{2q(r,\vartheta)}
(dr^2+r^2d\vartheta^2)+r^2\sin^2\vartheta d\varphi^2\bigr).$$
The constraint equation now has the form
$$\nabla^2\Phi=-f\Phi-2\pi\mu\Phi^5e^{2q},\eqno(7)$$
where
$$f={1\over 4}\biggl({\partial^2 q\over 
\partial r^2}+{1\over r}{\partial q\over\partial r}+
{1\over r^2}{\partial^2 q\over \partial \vartheta^2} \biggr).$$
Regularity of the metric on the axis $\vartheta=0$ 
implies the following boundary 
conditions for $\Phi$ and $q$ \cite{brill}:
$${\partial\Phi\over\partial\vartheta}(\vartheta=0,r) =0,$$
$${\partial q\over\partial\vartheta}(\vartheta=0,r) =0,$$
$$q(\vartheta=0,r)=0.$$
An identical set of conditions must hold 
for $\vartheta=\pi$. Asymptotically
we still set $\Phi\rightarrow A+
{B\over 2r}+O(1/r^2)$ and additionally
$q\rightarrow O(1/r^2)$.
Summarizing, if we have a function $q$, matter 
distribution $\mu$, constant $A$
and boundary condition on $\partial 
\Omega$ then $f$ and $\Phi$ are 
uniquely determined. Now the 
extensions are designated by $\widetilde M(q,\mu,A)$.
\par
We define a class of strongly 
nonconformally flat extensions in which
the influence of $q$ and source $f\Phi$ 
can be large enough to produce a closed
equipotential surface $\Phi=const$. 
\par
{\bf Definition.} Extension 
$\widetilde M(q,\mu,A)$ is {\bf strongly 
nonconformally flat} if there exist an equipotential 
surface $\Phi=const$ which encloses the 
support of function $f$ in such a way that: 
\par$\bullet$
volume $V$ inside $\Phi=const$ 
does not have common points with $\Omega$ (see picture),
\par$\bullet$
derivative in direction normal to this surface does not have
zero points: $n^i\partial_i\Phi|_{\partial V}\ne 0$. $n^i$ 
denotes normal unit vector.
\par
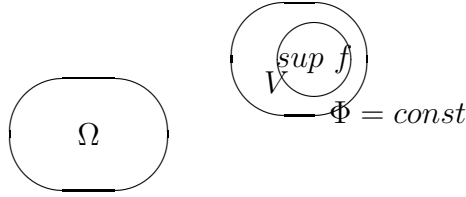
\unitlength=1mm
\special{em:linewidth 0.4pt}
\linethickness{0.4pt}
\begin{picture}(80,50)
\put(45,20){\oval(21,15)[]}
\put(45,20){\makebox(0,0)[cc]{$\Omega$}}
\put(75,30){\circle{10}}
\put(75,30){\makebox(0,0)[cc]{$sup~f$}}
\put(70,27){\makebox(0,0)[cc]{$V$}}
\put(73,30){\oval(18,15)[]}
\put(77,23){\makebox(0,0)[lc]{$\Phi=const$}}
\end{picture}
\par
The above two conditions intuitively mean that 
gravitational field  in the volume $V$
(where geometry is not 
conformally flat) is strong. It occurs that 
nonconformal extensions are more massive then conformally flat
vacuous ones.
\par
{\bf Theorem 2.} Assume that 
$\Omega$ and $\cal{PMW}$ are axially symmetric
and $\Omega$ is conformally flat. 
For any strongly nonconformally flat 
extension $\widetilde M(q_1,\mu_1,A_1)$ 
there exist a conformally flat
vacuous extension 
$\widetilde M(q=0,\mu=0,A_1)$ which has the ADM mass
not greater then mass of $\widetilde M(q_1,\mu_1,A_1)$.
\par
{\bf Proof.} Equation (7) may be rewritten in the form:
$$\nabla^2\log\Phi_1=
-f_1-2\pi\mu_1\Phi_1^4e^{2q_1}-(\nabla\log\Phi_1)^2.$$
Integrate the above formula over volume $V$:
$$\int\limits_{\partial V} 
{n^i\partial_i\Phi_1\over\Phi_1} d^2S=-
\int\limits_V [f_1+
2\pi\mu_1\Phi_1^4e^{2q_1}+(\nabla\log\Phi_1)^2]dV.$$
It can be easily checked that 
$\int\limits_{V}f_1dV=0$ (see \cite{brill}). 
The last two terms are nonnegative. From the other side we 
can use the condition that $\Phi_1$ 
is constant on $\partial V$. Hence
$$\int\limits_{\partial V}{n^i\partial_i 
\Phi_1\over \Phi_1}dS^2=
{1\over \Phi_1}\int\limits_{\partial V} 
n^i \partial_i\Phi_1 d^2S\leq 0.$$
Because $n^i\partial_i\Phi|_{\partial V}$ 
does not have zero points therefore
the last inequality implies that 
$n^i\partial_i\Phi|_{\partial V}\leq 0$
\par
Now we construct an extension  
$\widetilde M(q=0,\mu_2,A_1)$ which 
is conformally flat and has the same 
ADM mass as $\widetilde M(q_1,\mu_1,A_1)$.
Outside $V$ (in domain of conformal flatness) we 
choose $\widetilde M(q=0,\mu_2,A_1)$
in such a way that it is identical with original 
extension $\widetilde M(q_1,\mu_1,A_1)$ 
(this ensure the equality of ADM masses).
Inside $V$ we put $q=f=0$ and $\Phi=\Phi(\partial V)=const$.
Using eq.(7) one easy finds that condition 
$n^i\partial_i\Phi|_{\partial V}\leq 0$
implies nonnegativity of matter density on $\partial V$. 
The extension $\widetilde M(q=0,\mu_2,A_1)$
is conformally flat and has the same 
ADM mass as $\widetilde M(q_1,\mu_1,A_1)$.
Now we can utilize the theorem 1 
and conclude that there exist some
conformally flat vacuous extension which
is not more massive then $\widetilde M(q_1,\mu_1,A_1)$.
\par
Above theorems allows to conjecture that for conformally 
flat $\Omega$ vacuous conformally 
flat extensions are less massive
then other ones. Let us notice 
that conformal flatness may be 
intuitively understood as an 
absence of gravitational waves - 
a conformally flat system without matter is flat.
In this meaning conformally flat 
vacuous extensions do not contain
energy neither in the form of 
matter nor gravitational waves and
therefore are the least massive. 
\par
Conformally flat vacuous extensions are parameterized
by constant $A$ at infinity. It 
further stays to single out this particular
extension which has no horizons and minimalize the ADM mass.
\par
As an example we resolve this issue in 
the Schwarzschild geometry. Conformal
factor is defined by:
$$\Phi=1+{m_{Sch}\over 2r}.$$
We choose $\Omega$ to be a ball 
in a sense of background flat metric.
For simplifying the calculation 
assume that radius $\Delta$ of $\Omega$ 
is much smaller than the distance $R$
from the center of $\Omega$ 
to center of Schwarzschild solution.
We can translate the system of 
coordinates to the midpoint of $\Omega$.
$(\widetilde r,\widetilde\vartheta,\widetilde\varphi)$ 
are new spherical coordinates in which
the point $r=0$ refers to 
$\widetilde r=R$, $\widetilde\vartheta=0$.
In new coordinates
$$\Phi=1+{m_{Sch}\over 
2\sqrt{R^2+\widetilde r^2-2\widetilde rR\cos
\widetilde\vartheta}},\qquad \widetilde r\leq \Delta<R.$$
Let us now use the assumption that $\Delta\ll R$ 
and expand the above formula 
in power series for small $\widetilde r$.
$$\Phi=1+{m_{Sch}\over 2R}+{m_{Sch}\widetilde r\over 2R^2}
\cos\widetilde\vartheta\eqno(8)$$
Because of axial the symmetry 
and conformal flatness of $\Omega$ 
the assumptions of theorems are fulfilled. 
Hence we restrict ourselves to
extensions which are vacuous and conformally flat. Thus
for $\widetilde r>\Delta$ the hamiltonian 
constraint reduces to the Laplace 
equation. Hence
$$\Phi=A+{B\over 2\widetilde r}+\sum\limits_{l=1}^\infty{a_l 
P_l(\cos\widetilde\vartheta)\over \widetilde r^{l+1}},
\qquad \widetilde r>\Delta.
\eqno(9)$$
Comparing the last two equations on $\widetilde r=\Delta$ we get:
$$1+{m_{Sch}\over 2R}=A+{B\over 2\Delta},\eqno(10)$$ 
$$a_1={m_{Sch}\Delta^3\over 2R^2},\eqno(11)$$
$$a_{i>1}=0.\eqno(12)$$
Let us calculate the constant A from 
the first of the above formulas 
and insert it to eq. (3).
$$m_{ADM}=B\biggl(1+{m_{Sch}\over 2R}- 
{B\over 2\Delta} \biggr).\eqno(13)$$
Here we use parameterization by $B$ but 
it is equivalent (eq. (10)) with 
parameterization by $A$.
In order to evaluate Bartnik's 
mass function $m_B$ we have to determine 
constant $B$. Lower bound on $B$ ensues from 
the assumption of the nonnegativity
of matter density $\mu$. The only area 
where $\mu$ can be apriori negative 
is the boundary $\partial \Omega$ (interior of $\Omega$ 
and extension are vacuous). In generic 
case on $\partial\Omega$ we
obtain the shell type distribution of matter. From (2):
$$\mu=-{\nabla^2\Phi\over 2\pi\Phi^5}=
-{\Phi'_+-\Phi'_-\over 2\pi\Phi(\Delta)^5}
\delta(\widetilde r-\Delta),$$
where $\Phi'_-$ and $\Phi'_+$ are 
calculated from solution (8), (9) and the
condition (12).
$$\Phi'_-={\partial\Phi(\Delta_-)\over \partial\widetilde r}
={m_{Sch}\cos\widetilde\vartheta
\over 2R^2},$$
$$\Phi'_+={\partial\Phi(\Delta_+)\over 
\partial\widetilde r}=-{B\over 2\Delta^2}
-{2a_1\cos\widetilde\vartheta\over \Delta^3}.$$
Inserting the above formulas into equation 
for $\mu$ and using eq. (11) we get:
$$\mu={{3m_{Sch}\over 2R^2}
\cos\widetilde\vartheta +{B\over 2\Delta^2}
\over 2\pi \bigl( 1+
{m_{Sch}\over 2R}+{m_{Sch} \Delta\over 2R^2}
\cos\widetilde\vartheta\bigr)^5}
\delta(\widetilde r-\Delta).\eqno(14)$$
$\mu$ is everywhere nonnegative if
$$B\geq{3m_{Sch} \Delta^2\over R^2}=B_0.\eqno(15)$$
Upper bound for $B$ arises from 
the following reasoning. $B$ and $\mu$ are
constrained by the linear dependence (14). 
Because of the definition we assume that 
the analyzed extension does not contain horizons. We expect that
in extensions with horizon increasing 
of $\mu$ ({\bf than also} $B$, see eq. (14)) 
causes decreasing of the ADM 
mass (it can be checked on explicit example in 
spherical symmetry).  
Let us notice (eq.(13)) 
that such decreasing of mass occurs if
${B\over \Delta}>1+{m_{Sch}\over 2R}$. Therefore we put
$$B\leq \Delta\bigl(1+{m_{Sch}\over 2R}\bigr)=B_1.$$
One can easy find that
$$B_1\geq B_0$$
The ADM mass is minimalized for 
$B=B_0$ (see eq.(13)). Inserting $B_0$ (eq. (15)) to (13)
and omitting higher order terms in $\Delta/R$
we finally get
$$m_B=3m_{Sch}\bigl(1+{m_{Sch}\over 2R}\bigr)
\biggl({\Delta\over R}\biggr)^2.$$
Let us notice that the above formula does not 
allow to construct local 
energy density $m_B/{Volume}$ 
because such quantity diverges like $1/\Delta$ in each
point of a manifold. In other words 
binding energy of gravitational field is infinite.
\par
As a conclusion we would like to point
out that in wide class of
configurations the extension 
which minimize the ADM mass is vacuous. 
There was also presented the 
simple procedure of selecting extensions without 
horizons. These results 
suggests that in addition to mathematical elegance 
the definition seems also 
to be computable. The described approach can be 
applied (at least with some 
simple numerics) for any conformally flat volume 
$\Omega$.
\par
{\bf Acknowledgment:} I would like 
to thank Dr E.Malec for useful discussions
and reading the manuscript. This paper is supported by the polish 
government grant no {\bf 2PO3B 090 08}.

\eject
\par

\end{document}